\documentclass[aps,preprint,onecolumn,groupedaddress,superscriptaddress,showpacs]{revtex4-1}

\usepackage{graphicx}
\usepackage{color}
\usepackage{amsmath}
\usepackage{amsfonts}
\usepackage{amssymb}
\usepackage{dcolumn}
\usepackage{hyperref}
\usepackage{enumerate}
\usepackage{bbold}
\usepackage{cancel}
\usepackage{mathtools}
\usepackage{colonequals}
\usepackage{cleveref}
\usepackage{dsfont}
\Crefname{equation}{Eq.}{Eqs.}

\begin{document}
    \title{The Great Divide: Drivers of Polarization in the US Public}
\author{Lucas B\"ottcher}
\email{lucasb@ethz.ch}
\affiliation{Institute for Theoretical Physics, ETH Zurich, 8093 Zurich, Switzerland}
\affiliation{Center of Economic Research, ETH Zurich, 8092 Zurich, Switzerland}
\author{Hans Gersbach}
\affiliation{Center of Economic Research, ETH Zurich, 8092 Zurich, Switzerland}

\begin{abstract}
Many democratic societies have become more politically polarized, with the U.S.~as the main example. The origins of this phenomenon are still not well-understood and subject to debate. To better understand the mechanisms underlying political polarization, we develop a mathematical framework and employ information-theoretic concepts to analyze empirical data on political polarization that has been collected by Pew Research Center from 1994 to 2017. Our framework can capture the evolution of polarization in the Democratic- and Republican-leaning segments of the U.S.~public and allows us to identify its drivers.
Our findings provide empirical and quantitative evidence that political polarization in the U.S.~is mainly driven by strong and more left-leaning policy/cultural innovations in the Democratic party.
\end{abstract}

\date{\today}
\maketitle
\section{Introduction}
Political polarization is on the rise in many democratic societies~\cite{benkler2018network,gentzkow2016polarization,shapiro17,prior13,mccarty2016polarized,PewPolarization}, and yet the causes of this relatively recent development are not well-understood. In the U.S.~political polarization in terms of ideological distance between Republicans and Democrats has been growing significantly, so that it is now less likely to find a liberal Republican or a conservative Democrat~\cite{gentzkow2016polarization,PewPolarization}.
Several explanations for this finding have been put forward, including the increasing influence of new media and the Internet, the rising income inequality, elite polarization, and demographic changes~\cite{prior13,NBERw24462,Fiorina2008}. However, the growing use of the Internet, for instance, might not suffice to explain the observed polarization effects because polarization is largest among demographic groups that are least likely to use the Internet and social media~\cite{shapiro17}.

We focus on the basic forces of opinion formation and persuasion, and on how spreading of political and cultural ideas within populations that lean towards Democrats or Republicans can explain the evolution of political polarization, as observed in empirical data (see Fig.~\ref{fig:polarization_data}). In this context, mathematical models are able to offer insights into the dynamics of opinion formation, polarization, and related spreading processes~\cite{borghesi06,ugander2012structural,del2017modeling,krueger17,chuang2018age,dandekar2013biased,bottcher2018dynamical,bottcher2018clout,boettcher171,boettcher162,bottcher2017temporal,hoferer2019impact}. To quantify and characterize empirically observed polarization trends (see Fig.~\ref{fig:polarization_data}), we develop a mathematical framework of political change based on (i) individuals' diffusion from one ideological position to adjacent ones, and (ii) targeting of certain groups of individuals by \emph{influential actors} who spread their ideas to coalesce around political/cultural positions (henceforth simply called ``initiatives''). Influential actors are individuals or groups of individuals with a particular political or cultural interest and leaders of interest groups, movements, or political parties~\cite{Fiorina2008,muller2017populism}.

We apply Bayesian Markov chain Monte-Carlo and information-theoretic methods to quantitatively capture levels and changes of ideology distributions observed in the United States. We find that that a single parameter suffices to describe the evolution of polarization trends and we identify this polarization measure with the notion of \emph{initiative impact}. We use this measure to quantify relative changes in ideology distributions between Democratic- and Republican-leaning segments of U.S.~society. Our results suggest that the recent polarization in the U.S.~public is mainly driven by strong and more left-leaning policy/cultural concepts in the Democratic party. Prominent examples of such concepts are mandatory health insurance and various forms of identity politics~\cite{muller2017populism}.
\section{The model}
\label{sec:model}
\begin{figure*}
\centering
\includegraphics[width=0.49\textwidth]{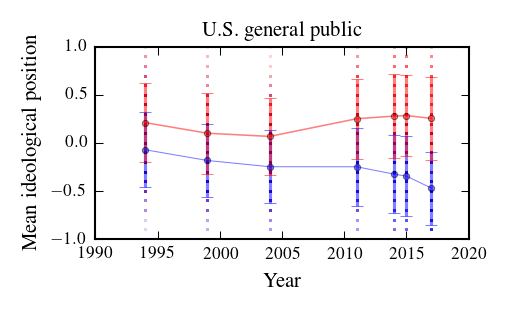}
\includegraphics[width=0.49\textwidth]{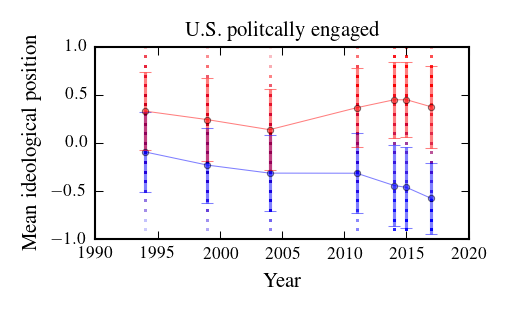}
  \caption{\textbf{Polarization in the U.S.~public.} We show the mean ideological position of the Democratic- (blue) and Republican-leaning (red) segments of U.S.~public from 1994 to 2017. Error bars indicate the observed standard deviation in each year. It is evident that polarization has been increasing in the past 15 years. The plotted data are based on a survey conducted by Pew Research Center (see Ref.~\cite{PewPolarization} for details).}
 \label{fig:polarization_data}
\end{figure*}
In this section, we first define a general and abstract Markov chain model to mathematically capture empirically observed polarization trends (see Fig.~\ref{fig:polarization_data}). We briefly describe the update dynamics and then focus on the characterization of the stationary distribution. 
\subsection{Definition of the ideology chain}
\label{sec:model_definition}
\begin{figure}
\centering
\includegraphics[width=0.7\textwidth]{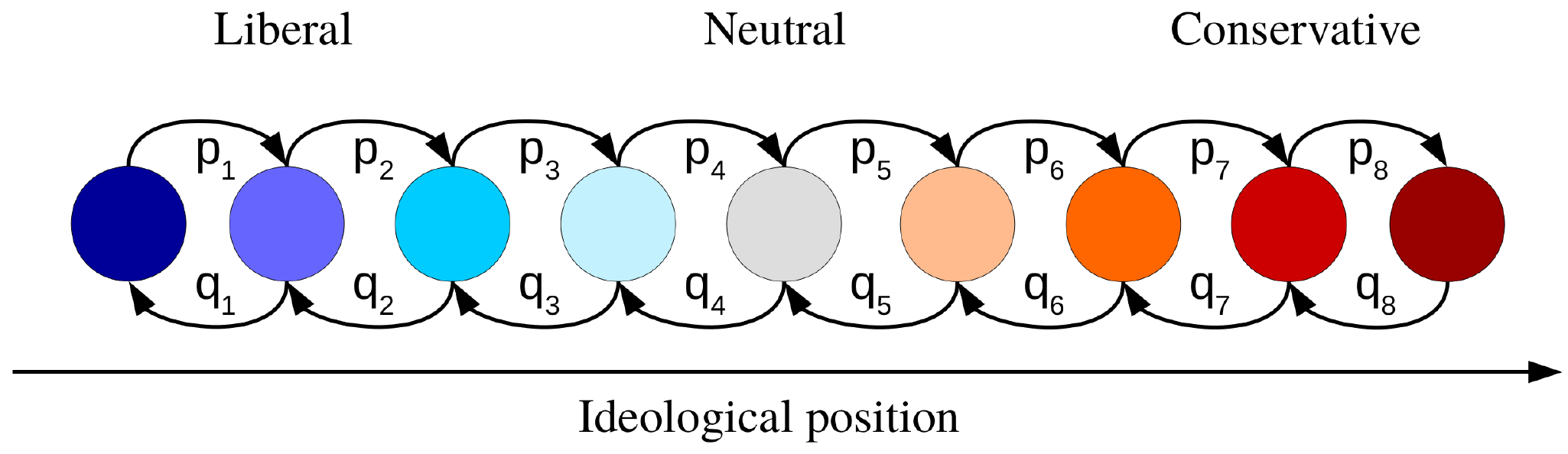}
  \caption{\textbf{Polarization model.} The political spectrum consists of $N$ different states and is divided in three groups: liberal, neutral, and conservative. For illustrative purposes, we set $N=9$ in this example. The transition probabilities are denoted by $\{p_i\}_{i\in \{1,\dots,9\}}$ and $\{q_i\}_{i\in \{1,\dots,9\}}$.} 
 \label{fig:model1}
\end{figure}

We proceed in three steps to mathematically describe initiatives and the diffusion of individuals from one ideological position to adjacent ones, with step 1 developed in Secs.~\ref{sec:model_definition} and \ref{sec:stationary}, step 2 in Sec.~\ref{sec:two_pop}, and step 3 in Sec.~\ref{sec:activists}. In the first step, we consider a one-dimensional chain which consists of $N$ different states denoted by $i$ ($i\in\{1,\dots,N\}$). $X_i$ denotes the fraction of the society in state $i$ and hence $\sum_{i=1}^{N} X_i = 1$. We use $x=2 (i-1)/(N-1)-1$ to map the index $i$ to an \emph{ideological position} $x\in[-1,1]$. These positions represent the political spectrum in the following way: Very liberal individuals are located at the beginning of the chain ($i=1$, $x=-1$), whereas strongly conservative ones are found at the opposite side ($i=N$, $x=1$). We next consider the evolution of a hypothetical society in discrete time. We interpret $X_i^n$ as the fraction of voters of type $i$ at time step $n\in\mathbb{N}$. For every $n$, it holds that
\begin{equation}
\sum_{i=1}^N X_i^n=1
\label{eq:normalization}
\end{equation}
as a normalization condition. We employ a simple model of social interactions and assume that individuals may change their ideological position through interactions with their ideological neighbors in the spirit of DeGroot's model~\cite{degroot1974reaching}, in which transitions from one state to another correspond to a social learning process in a group of communicating individuals. At the aggregate level, in a particular time step, we assume that transitions occur from state $i$ to its nearest neighbors ($i\rightarrow i+1$ and $i\rightarrow i-1$) with some probabilities $p_i\in(0,1)$ and $q_{i-1}\in(0,1-p_i)$. For the moment, these transition probabilities are taken as given and will be estimated later. At the boundaries of the opinion chain, the probability of becoming more ideologically extreme is zero ($q_0=p_N=0$). The probability of staying at a certain ideological position $i$ is given by $r_i=1-p_i-q_{i-1}$. These probabilities form the transition matrix $P$, with the following entries:
\begin{align}
P_{i i-1}=q_{i-1}, \quad P_{i i} = r_i, \quad \text{and} \quad P_{i i+1}=p_i.
\label{eq:transition_probs}
\end{align}
The probabilities in each row sum up to one, i.e.~$\sum_{j=0}^3 P_{i i-1}=1$. The aggregate-level behavior in our model can be traced back to individual behaviors in random matching and bounded confidence models~\cite{deffuant2000mixing,Hegselmann2002Opinion} in which individuals adopt sufficiently close opinions through communication (see Ref.~\cite{flache2011small} for a comprehensive account how such micro-level assumptions in a social network turn into macro-level implications and Ref.~\cite{horst2006equilibria} for a general theory about such social interactions). The probabilities $p_i$ and $q_i$ are then the resulting consequences at the aggregate level~\footnote{An alternative foundation of the model is competition of $N$ echo chambers. Individuals communicate in one echo chamber, but opinion leaders of echo chambers spread their views to adjacent echo chambers to increase their number of followers.}. We show an example of an ideology chain with $N=9$ states in Fig.~\ref{fig:model1}. For the sake of clarity, we do not include the self-loops described by $r_i$ in this figure. 
We next focus on the dynamics of the model to account for the diffusion of individuals from one ideology to adjacent ones. The initial values of all states are given by $X_i^{n=0}=X_i^0$. We refer to the row vector of all initial states as $X^0=\left(X_1^0,\dots,X_N^0\right)$. The time evolution of the ideology distribution is then described by $X^0 P^n  = X^n$.
\subsection{Stationary Distribution}
\label{sec:stationary}
We next determine the stationary ideology distribution. The update rule of state $X_i^n$ reads
\begin{equation}
X_i^{n+1}=(1-p_i-q_i)X_i^n+q_{i+1} X_{i+1}^n+p_{i-1} X_{i-1}^n.
\label{eq:update_states}
\end{equation}
We are not considering periodic boundaries, and thus find for state $i=1$,
\begin{equation}
X_1^{n+1}=(1-p_1)X_1^n+q_2 X_{2}^n.
\label{eq:update_left}
\end{equation}
As $n\rightarrow\infty$, we reach a stationary state. This implies that $X_i^{n+1}=X_i^n=X_i$ for all $i\in\{1,\dots,N\}$ and \Cref{eq:update_left} yields $X_{2}=\left(p_1/q_2\right) X_1$. Furthermore, based on \Cref{eq:update_states} we find by induction that~\cite{bottcher2019competing}
\begin{equation}
X_{i+1}=\left( \prod_{j=1}^i \frac{p_j}{q_{j+1}} \right) X_1.
\label{eq:solution_product} 
\end{equation}
To fulfill the normalization condition of \Cref{eq:normalization}, we set $X_1=1$ and divide each state $X_i$ by $\sum_{i=1}^N X_i$. The stationary distribution $X=(X_1,\dots,X_N)$ is unique since the transition matrix $P$ is irreducible and aperiodic~\cite{norris1998markov}. Irreducibility follows from the fact that any state in the Markov chain can be reached from any other state, and aperiodicity is satisfied because of $P_{ii}^n > 0$ for all $n\in\mathbb{N}$~\cite{norris1998markov}.

The data that we show in Fig.~\ref{fig:polarization_data} suggests that the ideology distributions of Democrats and Republicans are clearly distinguishable, but also exhibit stronger overlaps in the nineties. To capture both distributions with our model, we consider two opinion chains $A$ and $B$ in the subsequent sections, and account for the impact of influential actors and issue innovations in the third step.
\subsection{Two Populations}
\label{sec:two_pop}
\begin{figure*}[h]
\centering
\includegraphics[width=0.49\textwidth]{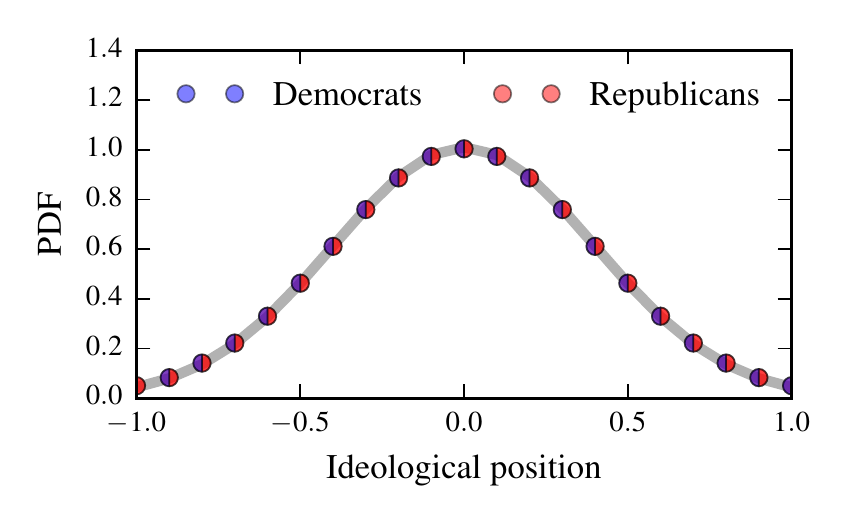}
\includegraphics[width=0.49\textwidth]{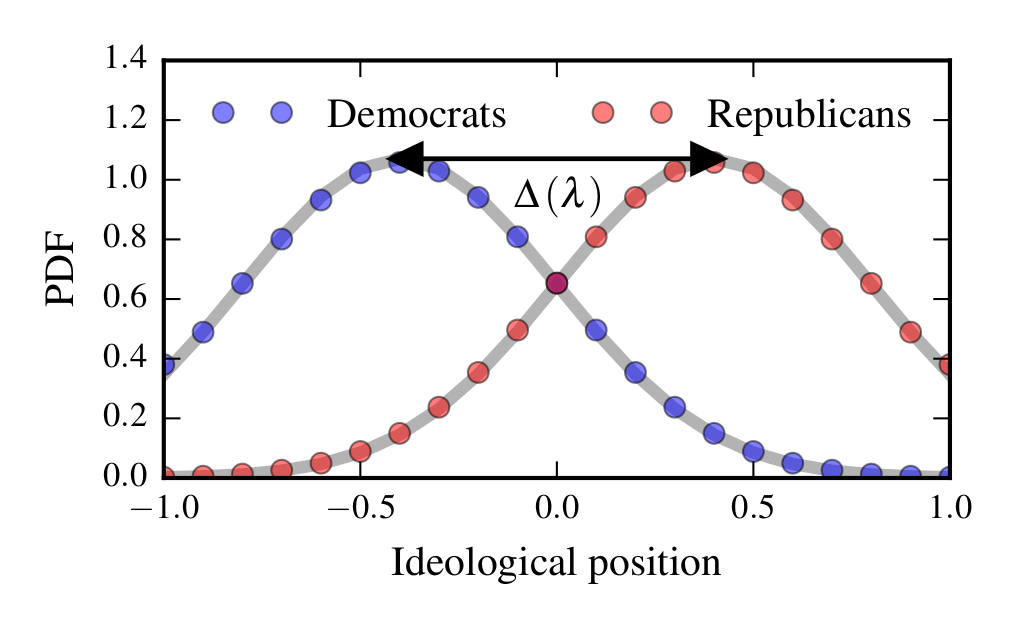}
\includegraphics[width=0.49\textwidth]{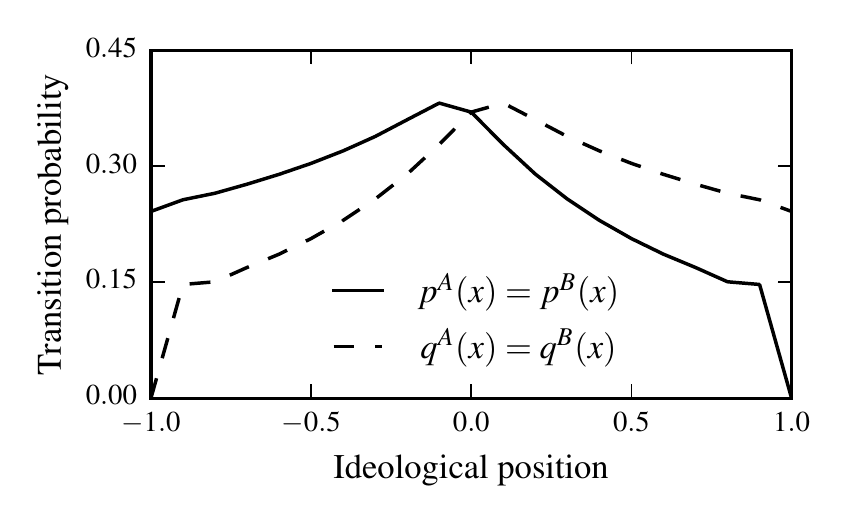}
\includegraphics[width=0.49\textwidth]{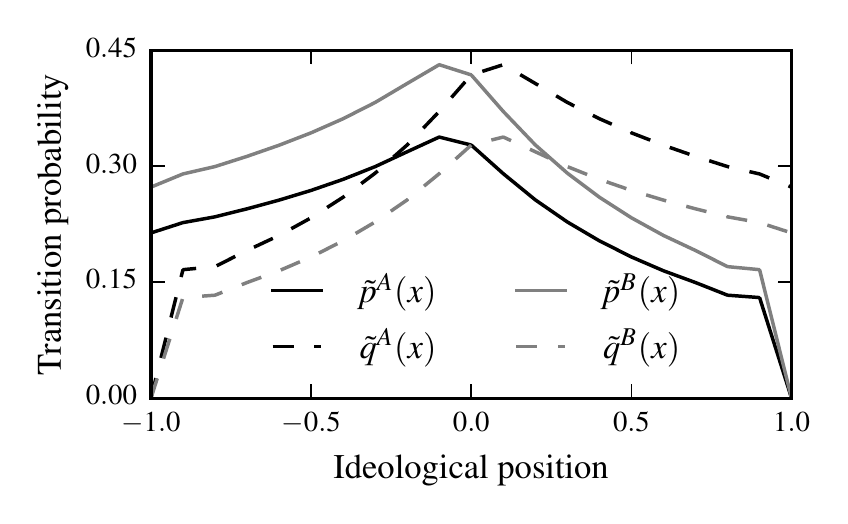}
 \caption{\textbf{The emergence of political polarization.} 
The top panels show an unpolarized (left panel) and a polarized (right panel) ideology distribution. Democrats are represented by blue dots and Republicans by red ones. The bottom panels show the corresponding transition probabilities which define the ideology distribution according to  Eqs.~(\ref{eq:solution_productA}) and (\ref{eq:solution_productB}) with $N=21$ states. We rescaled the probabilities in the right panel according to Eqs.~(\ref{eq:solution_productA_lambda}) and (\ref{eq:solution_productB_lambda}) by setting $\lambda=\lambda_A=\lambda_B=1.13$.} 
 \label{fig:onset}
\end{figure*}
In the second step, we introduce two populations in which members influence each other regarding their ideological position. This allows us to examine how the distribution of ideologies among Democrats and Republicans evolves over time.
Specifically, we consider two populations, $A$ and $B$, with the corresponding stationary ideology distributions given by \Cref{eq:solution_product}:
\begin{align}
X^A_{i+1}=\left( \prod_{j=1}^i \frac{p^A_j}{q^A_{j+1}} \right) X^A_1
\label{eq:solution_productA} 
\end{align}
and
\begin{align}
 X^B_{i+1}=\left( \prod_{j=1}^i \frac{p^B_j}{q^B_{j+1}} \right) X^B_1.
\label{eq:solution_productB} 
\end{align}
\subsection{Influential Actors}
\label{sec:activists}
In the third step, we introduce that influential actors of both parties transmit political/cultural concepts that increase the cohesion within each population and thus also increase the identity value of belonging to a population. The literature has identified the importance of such channels and corresponding influential actors (see e.g.~Refs.~\cite{Fiorina2008,muller2017populism}). In particular, we assume that influential actors inject concepts (initiatives) that increase the attractiveness of coalescing around ideological positions.
Mathematically, we account for this by describing the influence of such actors in the democratic and republican party as a rescaling of the transition probabilities of \Cref{eq:transition_probs} with $\lambda_A$ and $\lambda_B$ at a particular point in time. We interpret $\lambda_A$ and $\lambda_B$ as the initiative impact in each party. That is, we interpret $\lambda_A$ and $\lambda_B$ as the strength to coalesce around particular policy (or cultural) positions brought forward by influential actors. Typically, when initiatives and the related cultural identities are located to the left and right respectively, we would have $\lambda_A>1$ and $\lambda_B>1$. However, it is also possible that initiatives can be related to the center or even to the opposite side of the political spectrum which would imply that $\lambda_A$ and $\lambda_B$ can be smaller than one. This will turn out to be important in our empirical analysis. Specifically, the impact of influential actors on group $A$ (e.g., Democrats) at a particular point of time leads to
\begin{equation}
p_i^A\rightarrow p_i^A/\sqrt{\lambda_A} \quad \text{and} \quad q_i^A \rightarrow q_i^A \sqrt{\lambda_A}.
\label{eq:rescaling}
\end{equation}
For opinion group $B$ (e.g., Republicans), the rates are modified as follows:
\begin{equation}
p_i^B\rightarrow p_i^B \sqrt{\lambda_B} \quad \text{and} \quad q_i^B \rightarrow q_i^B /\sqrt{\lambda_B}.
\label{eq:rescalingB}
\end{equation}
Based on \Cref{eq:rescaling,eq:rescalingB}, we obtain the following modified stationary states:
\begin{equation}
X^A_{i+1}=\left( \prod_{j=1}^i \frac{p^A_j}{q^A_{j+1}} \right) \lambda_A^{-i} X^A_1
\label{eq:solution_productA_lambda} 
\end{equation}
and
\begin{equation}
X^B_{i+1}=\left( \prod_{j=1}^i \frac{p^B_j}{q^B_{j+1}} \right) \lambda_B^i X^B_1.
\label{eq:solution_productB_lambda} 
\end{equation}
In the following sections, we show that this approach is able to replicate the empirically-observed polarization.
In principal, we could also assume values of $\lambda_A$ and $\lambda_B$ that depend on the position in ideology space. It is, however, possible to capture a substantial part of the polarization effects with a constant value of $\lambda_A$ and $\lambda_B$, as shown in Sec.~\ref{sec:data_model}.
\section{Results}
\label{sec:results}
\begin{figure*}[h]
	\centering
	\includegraphics[width=0.49\textwidth]{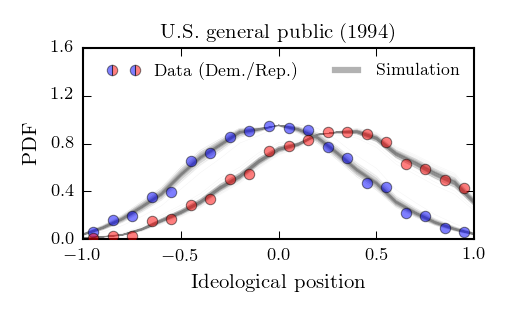}
	\includegraphics[width=0.49\textwidth]{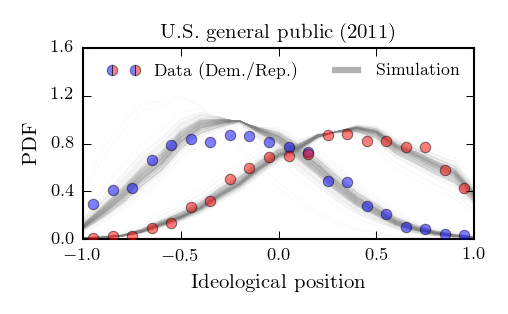}
	\includegraphics[width=0.49\textwidth]{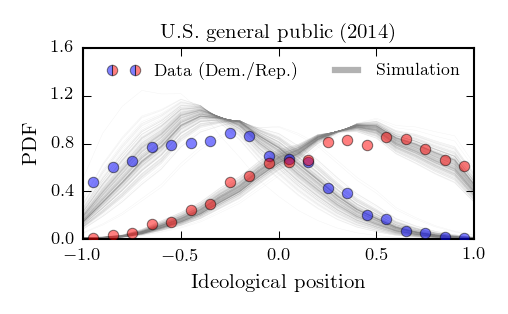}
	\includegraphics[width=0.49\textwidth]{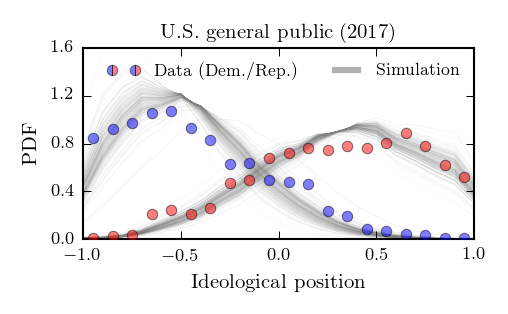}
	  \caption{
	  	\textbf{Polarization in the U.S.~general public in comparison with our model.} For the U.S.~general public, we show the ideology distributions for Democrats (blue dots) and Republicans (red dots). The dataset is based on a survey conducted by Pew Research Center (see Ref.~\cite{PewPolarization} for details). We initialized our model with the data of 1994 and only modified the distributions according to a transition probability rescaling as described by  Eqs.~(\ref{eq:solution_productA_lambda}) and (\ref{eq:solution_productB_lambda}).
  	 } 
	 \label{fig:model_vs_public}
\end{figure*}
We now focus on the applications and implications of the polarization model introduced in Sec.~\ref{sec:model}.
In Sec.~\ref{sec:analytical}, we discuss the onset of political polarization when influential actors in each party introduce new political ideas. We outline in Sec.~\ref{sec:data_model} that our mathematical framework is able to capture a substantial portion of the polarization effects which have been observed in the U.S.-American public in the past 25 years. Once initialized, only the two initiative impacts $\lambda_A$ and $\lambda_B$ are necessary to describe these polarization effects. We use a Bayesian Markov chain Monte Carlo approach to learn the parameter distributions of $\lambda_A$ and $\lambda_B$ from the empirical observations. Our results are consistent with a stronger polarization of the Democratic wing in the society compared to the Republican one.
\subsection{Emergence of political polarization}
\label{sec:analytical}
To study the emergence of political polarization in terms of our model as described in Sec.~\ref{sec:model}, we first consider an unpolarized society and then analyze the impact of influential actors. The ideology space is given by the interval $[-1,1]$. Moreover, we consider initially normally-distributed (unpolarized) ideologies with mean $\mu=0$ and variance $\sigma^2=0.16$, as illustrated in the upper left panel of Fig.~\ref{fig:onset}. We note that a distribution of ideologies within a party does not uniquely determine the transition probabilities for the diffusion of ideas (see SI). The reason is that, according to Eqs.~(\ref{eq:solution_productA}) and (\ref{eq:solution_productB}), only their fractions are relevant for the stationary distribution. To anchor meaningful transition probabilities, we take the fact into account that voters with polar ideological positions are less likely to undergo a transition to more moderate ideological positions. In addition, we would expect larger transition probabilities in the more neutral ideology regime. These two properties anchor the transition probabilities. We show an example of the corresponding transition probabilities $p^A(x)$, $q^A(x)$, $p^B(x)$, and $q^B(x)$ in the lower left panel of Fig.~\ref{fig:onset}. Also in the case of our empirical application in Sec.~\ref{sec:data_model}, the indeterminacy regarding the transition probabilities is resolved by taking the described effects into account. 

We now incorporate the impact of influential actors on voter ideologies as described by  Eqs.~(\ref{eq:solution_productA_lambda}) and (\ref{eq:solution_productB_lambda}), and rescale the transition probabilities accordingly to obtain
\begin{align}
\tilde{p}^A(x) = p^A(x)/\sqrt{\lambda}, \quad \tilde{q}^A(x) = q^A(x)\sqrt{\lambda},
\label{eq:rescaling2}
\end{align}
and
\begin{align}
\tilde{p}^B(x) = p^B(x) \sqrt{\lambda}, \quad \tilde{q}^B(x) = q^B(x)/\sqrt{\lambda},
\label{eq:rescaling3}
\end{align}  
where we assumed $\lambda=\lambda_A=\lambda_B$.
In the upper right panel of Fig.~\ref{fig:onset}, we show that a rescaling with $\lambda$ leads to shifted normal distribution with mean $\mu(\lambda)$ and an invariant variance. In this example, we set $\lambda=1.13$. This finding suggests that a rescaling of transition probabilities according to Eqs.~(\ref{eq:rescaling2}) and (\ref{eq:rescaling3}) leads to a polarized society.
The corresponding transition probabilities and their rescaled versions are shown in the lower right panel of Fig.~\ref{fig:onset}. As shown in the upper right panel of Fig.~\ref{fig:onset}, it is possible to quantify polarization in terms of $\Delta(\lambda)$ which is $2 \mu(\lambda)$ in this example. Higher values of $\lambda$ lead to an increase in $\Delta(\lambda)$. In other words, polarization is monotonically increasing with initiative impact.
\subsection{Polarization in the American public}
\label{sec:data_model}
\begin{figure*}[h]
	\centering
	\includegraphics[width=0.49\textwidth]{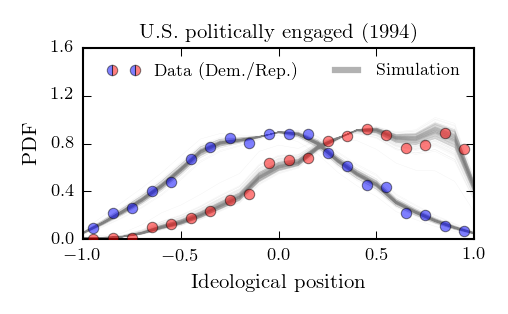}
	\includegraphics[width=0.49\textwidth]{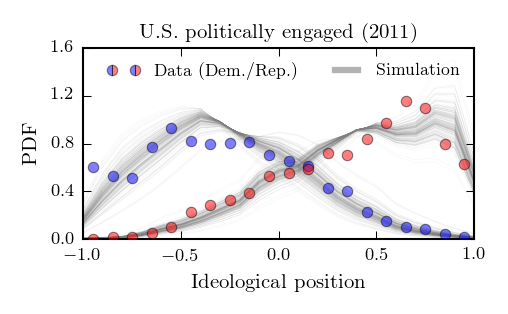}
	\includegraphics[width=0.49\textwidth]{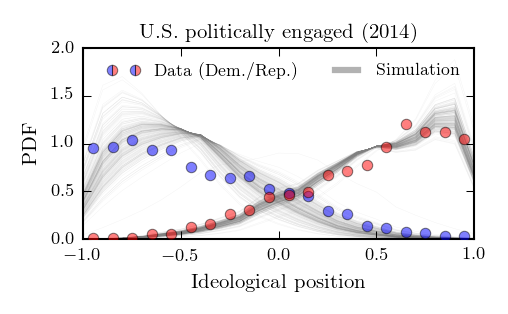}
	\includegraphics[width=0.49\textwidth]{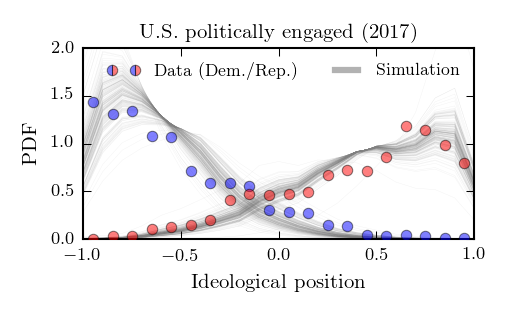}
	  \caption{
	  	\textbf{Polarization in the U.S.~politically-engaged public in comparison with our model.} For the U.S.~politically engaged public, we show the ideology distributions for Democrats (blue dots) and Republicans (red dots). The dataset is based on a survey conducted by Pew Research Center (see Ref.~\cite{PewPolarization} for details). We initialized our model with the data of 1994 and only modified the distributions according to a transition probability rescaling as described by Eqs.~(\ref{eq:solution_productA_lambda}) and (\ref{eq:solution_productB_lambda}).
  	 } 
	 \label{fig:model_vs_engaged}
\end{figure*}
After having outlined the basic polarization mechanism in our model, we now focus on the evolution of polarization as observed in empirical data on ideology distributions of US citizens from 1994 to 2017. The dataset is based on a Pew Research Center survey~\cite{PewPolarization}. For a detailed overview about the survey methodology, see Ref.~\cite{PewSurvey}. Ideology distributions are available for the Democratic- and Republican-leaning segments of the U.S.~general and politically-engaged public. We initialize our opinion chain by using the empirical ideology distributions of 1994 to determine the transition probabilities as defined by \Cref{eq:transition_probs} with a maximum-likelihood estimation. As in Sec.~\ref{sec:analytical}, we consider the case where the transition probabilities are monotonically increasing towards the center. After the initialization procedure, the parameters $\lambda_A$ and $\lambda_B$ are the only two free parameters in our model. They will describe the observed ideology distribution according to Eqs.~(\ref{eq:solution_productA_lambda}) and (\ref{eq:solution_productB_lambda}). In the next step, we use a Bayesian Markov chain Monte-Carlo approach to learn the distributions of both parameters that best describe our data~\cite{pandey2013comparing,gelman2013bayesian}. The theoretical background is presented in the SI.

After having initialized our model with a maximum-likelihood estimate with respect to the empirical data of 1994, we perform a Bayesian Markov chain Monte-Carlo parameter estimation for $\lambda_A$ and $\lambda_B$. The data and the corresponding estimates of the ideology distributions are shown in Figs.~\ref{fig:model_vs_public} and \ref{fig:model_vs_engaged}. For 1994, the ideology distributions of Democrats and Republicans are almost Gaussian and centered around the origin. In subsequent years, polarization becomes more and more apparent. The division is much more radical in the case of the politically-engaged citizens, compared to the general public. The plots also reveal that, after the initial transition probability estimation, our two-parameter model is able to describe the time evolution of the two ideology distributions quite well. Hence, a substantial portion of the complex ideology and identity formation process can be captured by a multiplicative rescaling of the transition probabilities according to \Cref{eq:rescaling2}.
\begin{figure*}[h]
	\centering
	\includegraphics[width=0.49\textwidth]{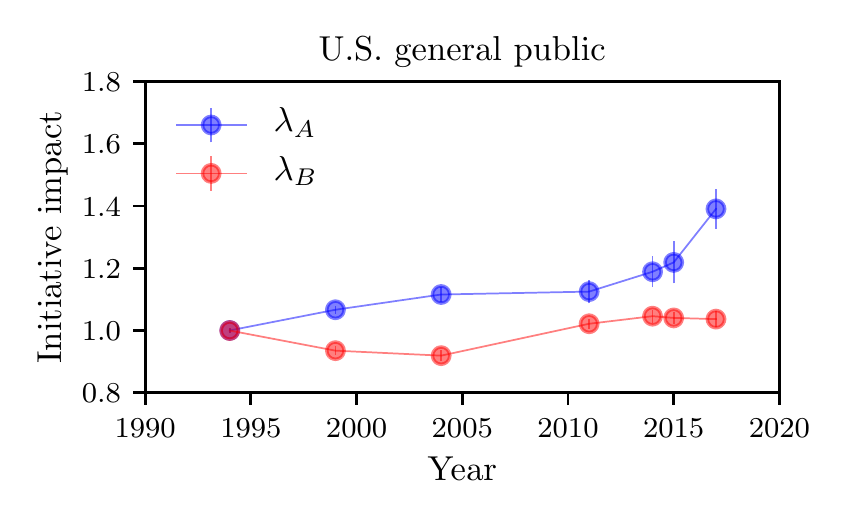}
	\includegraphics[width=0.49\textwidth]{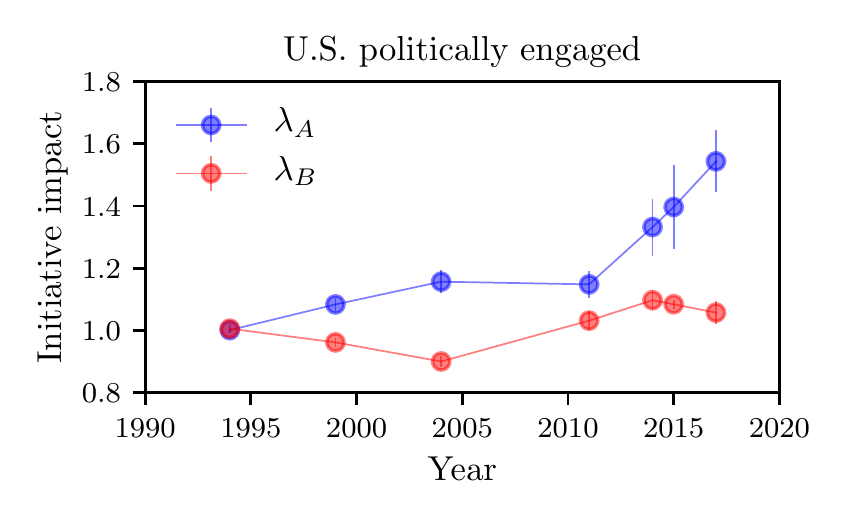}
	\includegraphics[width=0.49\textwidth]{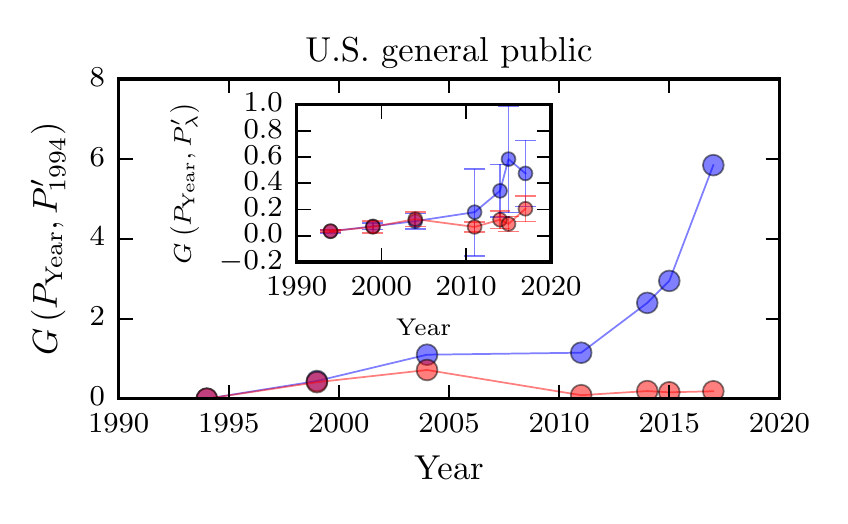}
	\includegraphics[width=0.49\textwidth]{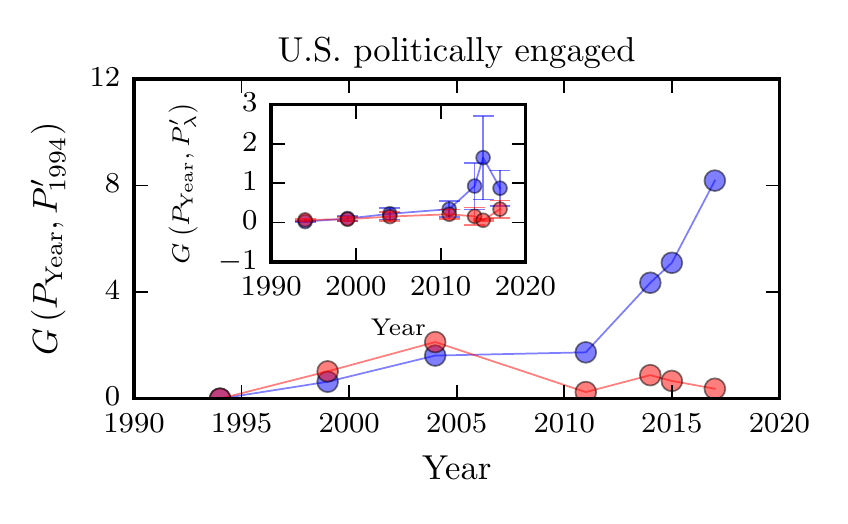}
	  \caption{
	  	\textbf{Initiative impacts and relative entropy over different years.} In the upper panels, we estimated the initiative impact $\lambda_A$ and $\lambda_B$ as described by \Cref{eq:solution_productA_lambda,eq:solution_productB_lambda}. We see that the polarization effects in the case of self-identified Democrats is larger than in the case of their republican counterparts. In addition, we compute the relative entropy according to \Cref{eq:kullback_leibler} and show the results in the lower panels. $P_{\text{Year}}$, $P^\prime_{1994}$, and $P^\prime_{\lambda}$ are the distributions of the data, of the data in 1994, and of our model, respectively. Blue dots represent Democrats and red ones Republicans.
  	 } 
	 \label{fig:lambdaA_lambdaB}
\end{figure*}

For every year, we show the corresponding distributions of $\lambda_A$ and $\lambda_B$ in the SI in Fig.~S2. In addition, we illustrate the time evolution of $\lambda_A$ and $\lambda_B$ in Fig.~\ref{fig:lambdaA_lambdaB}. We interpret the initiative impact $\lambda$ as a polarization measure. This allows us to analyze the polarization dynamics more systematically. The data presented in the upper panels of Figs.~\ref{fig:lambdaA_lambdaB} and S2 make clear that the democratic initiative impact $\lambda_A$ increases substantially over time. The distributions of $\lambda_A$ are also getting broader. This is a consequence of additional effects which we cannot describe by rescaling the initially-determined transition probabilities. Interestingly, the polarization behavior associated with the republican initiative impact $\lambda_B$ differs significantly from the democratic one. In fact, in 1999 and 2004, the values of $\lambda_B<1$ imply that the Republicans are ideologically moving towards the center. Just after 2011, the values of $\lambda_B>1$ suggest that more right-wing initiatives have been transmitted. As a result, Republicans are moving to the right in the political spectrum.
Our results clearly suggest, however, that polarization is mainly driven by issue forces acting on the Democratic wing of society.

In the last step, we compare the difference of the ideology distributions in each year in terms of the of the relative entropy (or Kullback-Leibler divergence)
\begin{equation}
G\left(P_{\text{Year}},P^\prime\right)=\sum_x P_{\text{Year}}(x) \ln\left[\frac{P_{\text{Year}}(x)}{P^\prime(x)}\right],
\label{eq:kullback_leibler}
\end{equation}
where $P_{\text{Year}}$ is the distribution of the empirical data in the respective year, and $P^\prime$ the one of the model, or of another reference. 
We first determine the relative entropy of $P_{\text{Year}}$ with respect to the empirical data distribution $P^\prime_{1994}$ of 1994. The result is shown in the lower panels of Fig.~\ref{fig:lambdaA_lambdaB}. We again see that the democratic distributions deviate much more from the ones of 1994 compared to the republican distributions---another indicator of a larger polarization effect in the Democratic wing of the society. The insets in the lower panels of Fig.~\ref{fig:lambdaA_lambdaB} show the relative entropy of $P_{\text{Year}}$ with respect to the distribution of our model $P^\prime_\lambda$. The small values of $G\left(P_{\text{Year}},P^\prime_\lambda\right)$ indicate that our model captures the empirical distributions well.
\section{Conclusion}
We developed a mathematical framework and used information-theoretic methods to analyze empirical ideology distributions of the Democratic- and Republican-leaning segments of the U.S.-American public. Our framework is based on two forces: varying strength of initiatives in each party, and the corresponding diffusion of these concepts in society. The time evolution of the observed ideology distributions is well-captured by only two parameters, namely the two initiative impacts. We use these parameters as polarization measures because of their ability to describe the increasing gap between the ideology distributions. Our analysis provides empirical and quantitative evidence that strong initiatives in the Democratic party have been the main drivers of the great divide that emerged in recent decades between the Democratic- and Republican-leaning population. Future studies may apply the proposed framework to quantify polarization trends in other countries.
\section*{Data availability}
The data that support the findings of this study are openly available from Ref.~\cite{PewPolarization}.
\section*{Code availability}
The codes that were used to perform the numerical analyses are available from the corresponding author upon request.
\begin{acknowledgments}
We thank participants at various workshops for helpful comments. Pew Research Center bears no responsibility for the analyses or interpretations of the data presented here. The opinions expressed herein, including any implications for policy, are those of the authors and not of Pew Research Center.
\end{acknowledgments}
%
%
%
%
%
\newpage
\appendix
\setcounter{figure}{0}    
\renewcommand{\thefigure}{S\arabic{figure}}
\begin{center}
  { \Large{SUPPORTING INFORMATION (SI)}}
\end{center}
\begin{figure}
	\centering
	\includegraphics[width=0.49\textwidth]{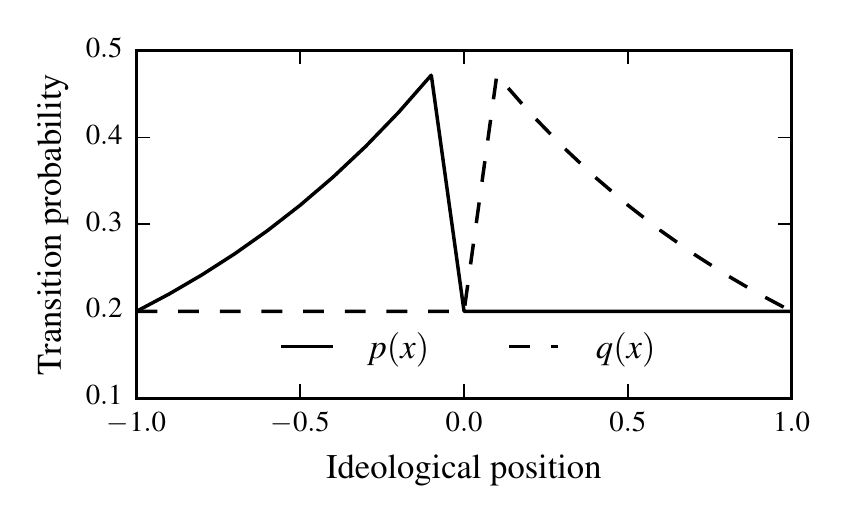}
	\includegraphics[width=0.49\textwidth]{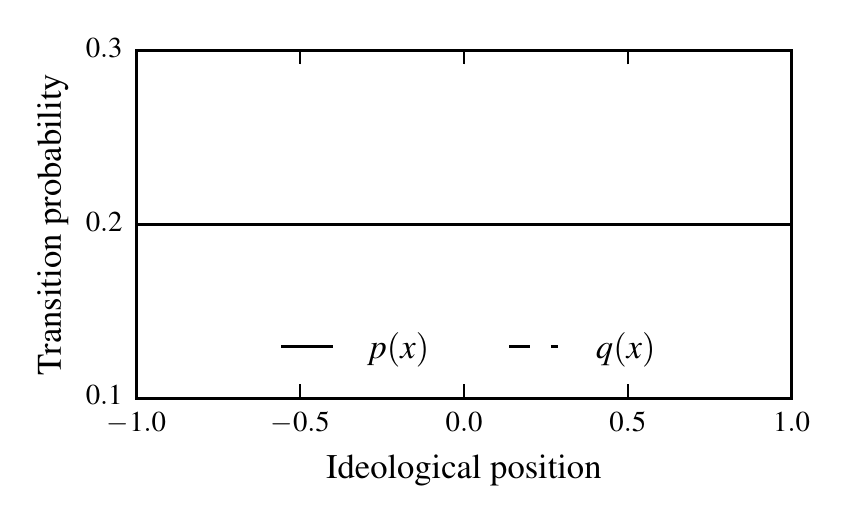}
	\includegraphics[width=0.49\textwidth]{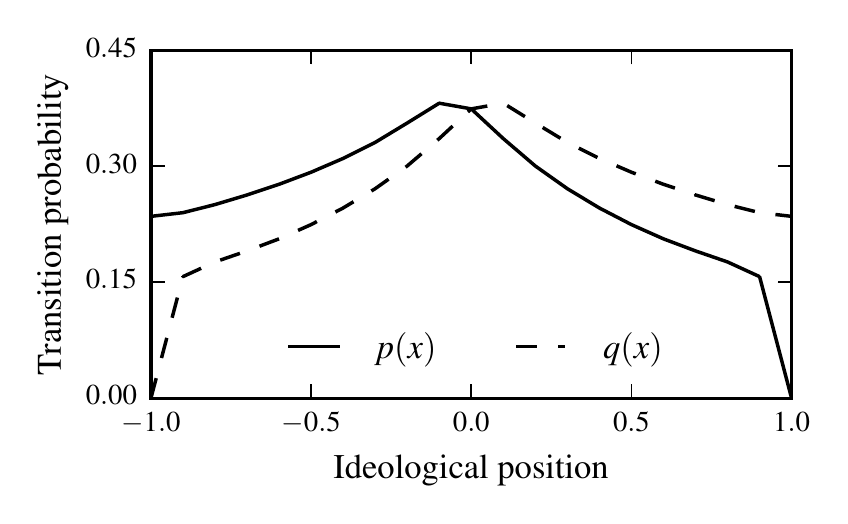}
	\includegraphics[width=0.49\textwidth]{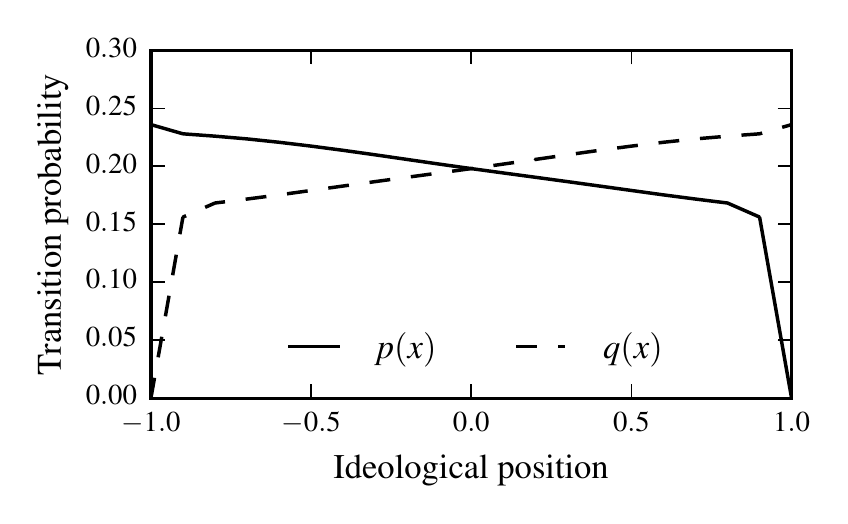}
	\includegraphics[width=0.49\textwidth]{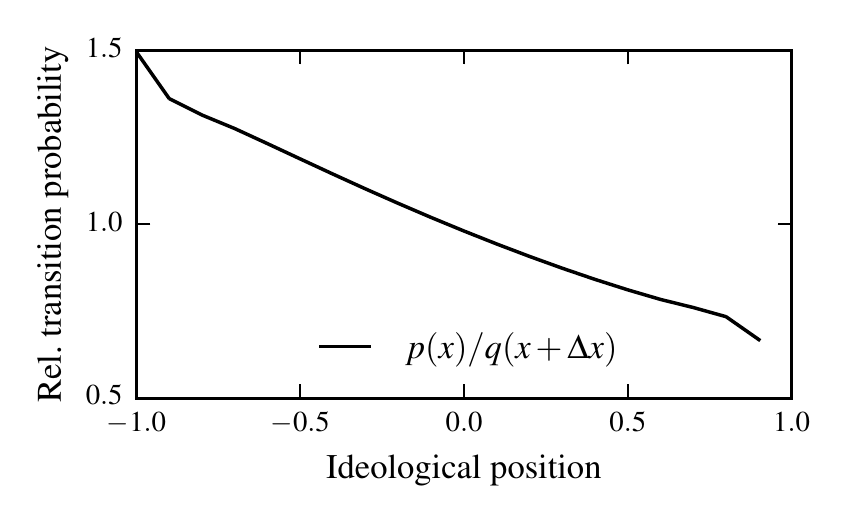}
	\includegraphics[width=0.49\textwidth]{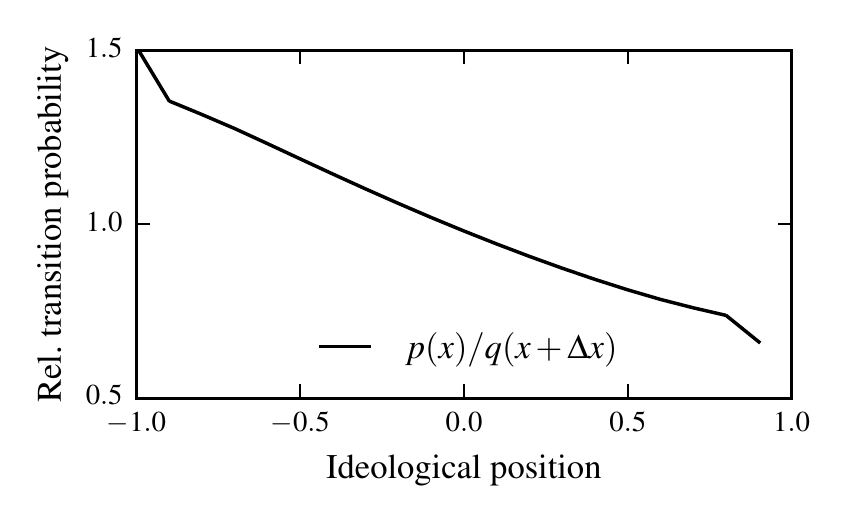}
	  \caption{
	  	\textbf{Determining transition probabilities.} We numerically determine the transition probabilities describing the Gaussian ideology distribution shown in Fig.~3 of the main text. In the upper left panel, we start from initial transition probabilities as defined by Eqs.~\eqref{eq:initial_1} and \eqref{eq:initial_2} to apply a least-square optimization method to determine the transition probabilities of the Gaussian ideology distribution. The results are shown in the middle left panel and the corresponding fraction $p(x)/q(x+\Delta x)$ is shown in the lower left panel. In the right panels, we repeat the same procedure by assuming an initially uniform probability distribution. The fractions $p(x)/q(x+\Delta x)$ are unaffected by this choice.
  	 } 
	 \label{fig:trans_probs}
\end{figure}
\section{Determining Transition Probabilities}
\label{app:trans_probs}
According to Eqs.~(10) and (11) of the main text, the transition probabilities $p^A_i$, $q^A_i$, $p^B_i$, and $q^B_i$ of the discrete locations $i\in\{1,\dots,N\}$ are not uniquely determined by a given ideology distribution. The reason is that only the fractions $p_i/q_{i+1}$ are relevant for determining the stationary distribution of the ideology chain. To identify meaningful transition probabilities, we account for the fact that voters with polar ideological position are less likely to shift towards more moderate ideological positions. Furthermore, we expect larger transition probabilities in the more neutral ideology regime. As initial values, we consider the transition probabilities
\begin{equation}
p_i=
\begin{cases}
0.2 \cdot \alpha^i &\text{for}~i< N/2,\\
0.2 &\text{otherwise},
\end{cases}
\label{eq:initial_1}
\end{equation}
and
\begin{equation}
q_i=
\begin{cases}
0.2 \cdot \alpha^i &\text{for}~i> N/2,\\
0.2 &\text{otherwise}.
\end{cases}
\label{eq:initial_2}
\end{equation}
We consider $N=20$ states and we set $\alpha=1.1$ as starting values. The corresponding probability curves for the chosen starting values are shown in the upper left panel of Fig.~S\ref{fig:trans_probs}. We discuss the influence of noise on transition probabilities and the resulting opinion distributions in Sec.~\ref{sec:noise}. After this initialization, we apply a least-square optimization method to obtain the transition probabilities that describe the Gaussian ideology distribution shown in Fig.~3 of the main text. The corresponding transition probabilities and fractions $p_i/q_{i+1}$ are shown in the middle left and lower left panel of Fig.~S\ref{fig:trans_probs}, respectively. In the right panels of Fig.~S\ref{fig:trans_probs}, we show that starting from another initial distribution may  lead to different transition probabilities, but to the same fractions $p_i/q_{i+1}$. We note that according to Eqs.~(10) and (11) of the main text, a given ideology distribution is uniquely determined by the fractions $p_i/q_{i+1}$.
\newpage
\section{Initiative Strength Distributions}
\label{app:distr}
\begin{figure}
	\centering
	\includegraphics[width=0.49\textwidth]{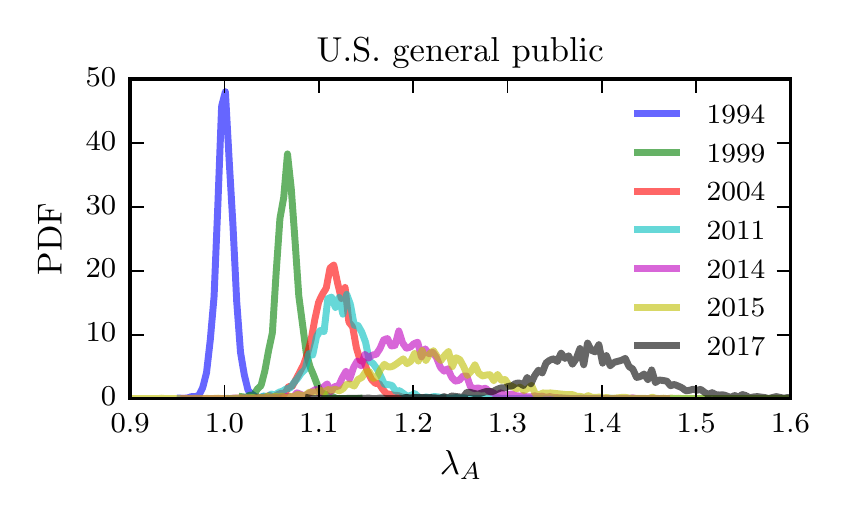}
	\includegraphics[width=0.49\textwidth]{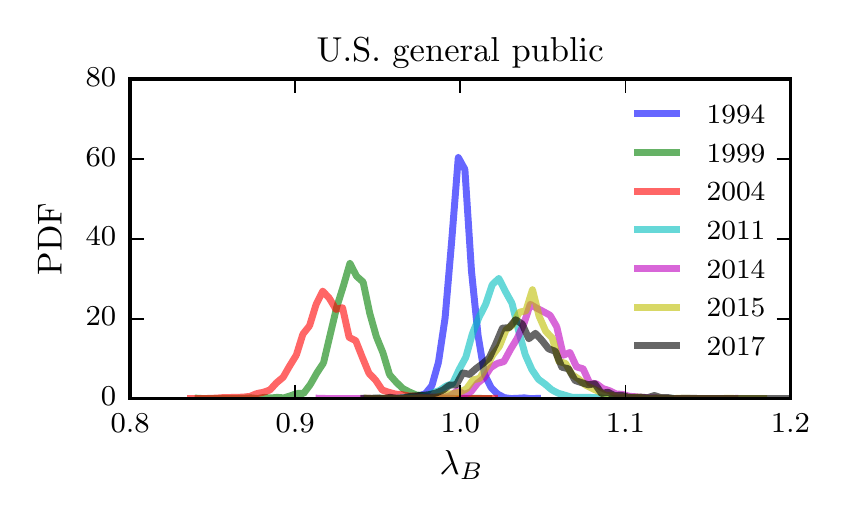}
	\includegraphics[width=0.49\textwidth]{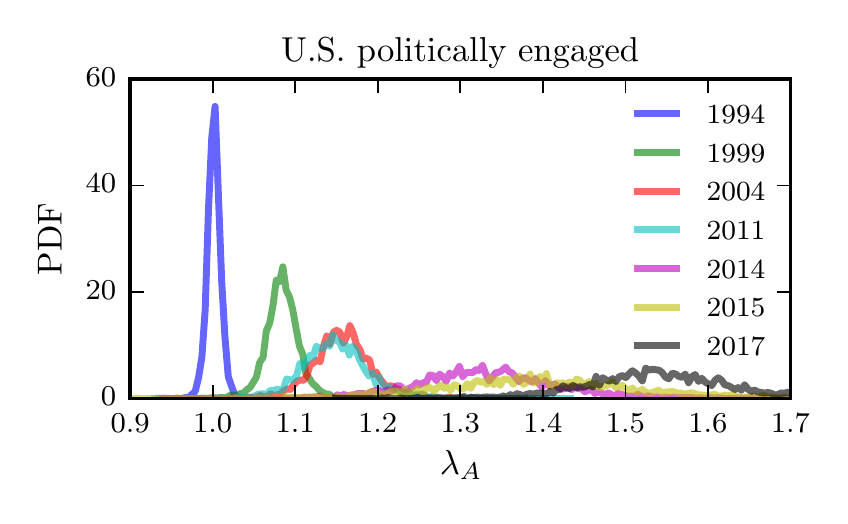}
	\includegraphics[width=0.49\textwidth]{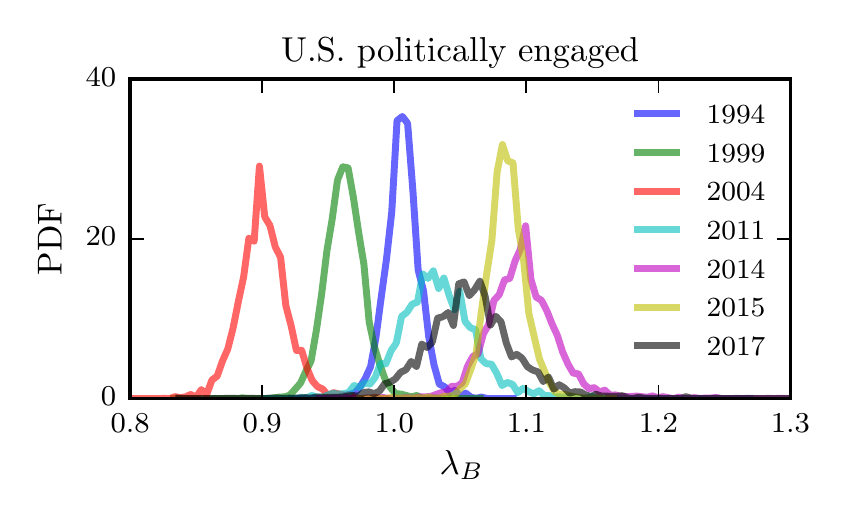}
	  \caption{
	  	\textbf{Initiative strength distributions across time.} We determine the probability density functions (PDFs) of the initiative strengths $\lambda_A$ and $\lambda_B$ using Bayesian Markov Chain Monte-Carlo. The values of $\lambda_A$ become larger over time, whereas the values of $\lambda_B$ are more concentrated around unity.
  	 } 
	 \label{fig:lambdaA_lambdaB_distr}
\end{figure}
In the main text we have outlined that the initiative strength $\lambda_A$ of self-identified democrats grew in the period from 1994 to 2017. In contrast, the initiative strength $\lambda_B$ of self-identified Republicans remained much more concentrated around unity and grew only little in the last years. In Fig.~6 of the main text, we show the mean values and standard deviations of $\lambda_A$ and $\lambda_B$. These quantities were obtained from the corresponding Bayesian Markov Chain Monte-Carlo distributions of Fig.~S\ref{fig:lambdaA_lambdaB_distr}. These distributions also show that the initiative strength $\lambda_A$ increases with time. In addition, we observe the broadening of the distributions that is also described by the increasing standard deviation in Fig.~6 of the main text. This originates from the fact that not all details of the evolution of the ideology distribution can be captured by the evolution of a single-parameter which we referred to as \emph{initiative strength}. In the case of self-identified Republicans, the corresponding distributions of $\lambda_B$ are more concentrated around the initial distribution of 1994. In addition, the broadening is also less pronounced.
\section{Bayesian Markov chain Monte Carlo}
\label{app:app_bayesian}
To compute the distribution of the parameter set $\theta\in\{\lambda_A,\lambda_B\}$ of the ideology chain as defined by Eqs.~(10) and (11) in the main text, we have to determine the probability distribution $P\left(\theta|D\right)$ given our data $D$ on the distribution of ideologies of self-identified Democrats and Republicans. According to Bayes' theorem, we express the posterior distribution
\begin{equation}
P\left(\theta|D\right)\propto P\left(D|\theta\right)P\left(\theta\right),
\label{eq:bayes}
\end{equation}
in terms of the likelihood function $P\left(D|\theta\right)$ and the prior parameter distribution $P\left(\theta\right)$. We assume that the likelihood function $P\left(D|\theta\right)$ is a Gaussian distribution
\begin{equation}
P\left(D|\theta\right) \propto \exp\left(-\frac{E^2}{2 \sigma^2} \right)
\label{eq:likelihood}
\end{equation}
of the error $E$ with zero mean and variance $\sigma^2$.
We use the prediction of our model $X_i(\theta)$ for a given parameter set $\theta$ and the actual ideology data $D_i$ ($i\in\{1,2,\dots,20\}$) to compute the least square error
\begin{equation}
E^2=\sum_{i=1}^{20} \left[D_i - X_i\left(\theta\right)\right]^2
\label{eq:error}
\end{equation}
as our error estimate in Eq.~\eqref{eq:likelihood}~\cite{pandey2013comparing}. For a given prior parameter distribution $P\left(\theta \right)$, we compute the posterior distribution $P\left(D|\theta\right)$ using Bayesian
Markov chain Monte-Carlo sampling with a Metropolis update scheme~\cite{pandey2013comparing,gelman2013bayesian}. After initializing the parameter vector with $\theta^0$ drawn from the prior distribution $P\left(\theta\right)$, the $n$th iteration of the algorithm is defined as follows:
\begin{enumerate}
\item A new parameter set $\theta^\ast$ is drawn from the proposal distribution $J\left(\theta^\ast | \theta^{n} \right)$.
\item The acceptance probability for $\theta^\ast$ is computed according to (Metropolis algorithm)
\begin{equation}
r=\min\left(\frac{P\left(\theta^\ast|D\right)}{P\left(\theta^n|D\right)},1\right)=\min\left(\frac{P\left(D|\theta^\ast\right)P\left(\theta^\ast\right)}{P\left(D|\theta^n\right)P\left(\theta^n\right)},1\right).
\label{eq:metropolis_app}
\end{equation}
\item Draw a random number $\epsilon\sim U(0,1)$ and set
\begin{equation}
\theta^{n+1} = \begin{cases} \theta^\ast &\text{if } \epsilon < r, \\
\theta^{n} & \text{otherwise.} \end{cases} 
\end{equation}
\end{enumerate}
This update procedure implies that a new parameter set is always accepted if the new likelihood function value is greater than or equal to the one of the previous iteration. That is, if $P\left(D|\theta^\ast\right)\geq P\left(D|\theta^{n}\right)$. For the described Metropolis algorithm, the proposal distribution $J\left(\theta^\ast | \theta^{n} \right)$ must be symmetric. According to Ref.~\cite{gelman2013bayesian}, a multivariate Gaussian proposal distribution
\begin{equation}
J\left(\theta^\ast | \theta^{n} \right) \sim N\left(\theta^{n} | \tilde{\lambda}^2 \Sigma \right)
\end{equation}
may be used as proposal distribution. The covariance matrix is denoted by $\Sigma$ and the corresponding scaling factor by $\tilde{\lambda}$. Every 500 iterations, the covariance matrix and the scaling factor are updated, using the following update procedure~\cite{pandey2013comparing,gelman2013bayesian}:
\begin{align}
\begin{split}
\Sigma_{k+1} &= p \Sigma_{k}+(1-p) \Sigma ^\ast, \\
\tilde{\lambda}_{k+1} &= \tilde{\lambda}_{k}\exp\left(\frac{\alpha^\ast-\hat{\alpha}}{k}\right),
\end{split}
\end{align}
where $\Sigma^\ast$ and $\alpha^\ast$ are the covariance matrix and the acceptance rate of the last 500 iterations, respectively. The remaining parameters are $p=0.25$, $\Sigma_0=\mathds{I}$, and $\lambda_0=2.4/\sqrt{d}$, where $\mathds{I}$ is the $d\times d$ identity matrix and $d$ the number of estimated parameters. The target acceptance rate is~\cite{gelman2013bayesian}
\begin{equation}
\hat{\alpha}= \begin{cases}0.44 &\text{if }d=1, \\
0.23& \text{otherwise.} \end{cases} 
\end{equation}

\begin{figure}[h]
\begin{minipage}{0.49\textwidth}
\centering
\includegraphics{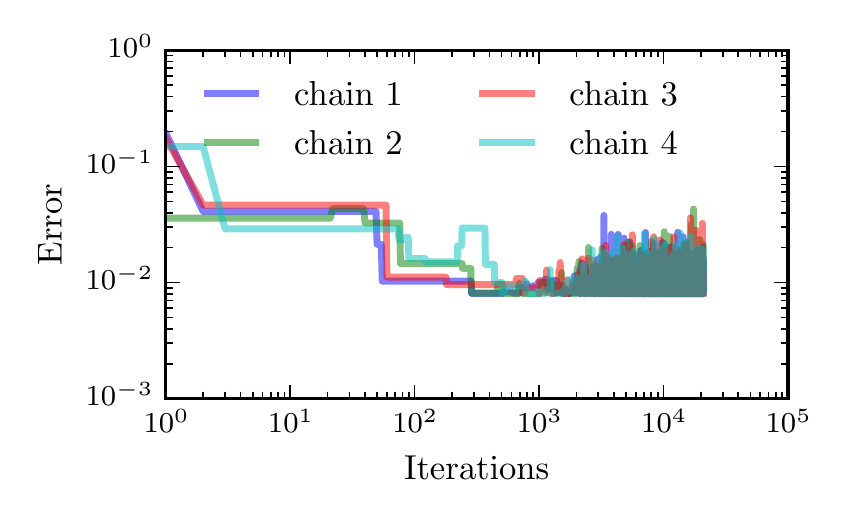}
\end{minipage}
\caption{\textbf{Convergence of Bayesian
Markov Chain Monte Carlo Parameter Estimation.} We show the evolution of the error as defined by Eq.~\eqref{eq:error} for four different chains with different initial conditions. The algorithm is considered to have converged to the posterior distribution if the variance \emph{between} the chains is similar to the variance \emph{within} the chains (which is precisely captured by the Gelman-Rubin Test)~\cite{gelman2013bayesian}. We first let the system evolve for $10^4$ iterations to then apply the Gelman-Rubin Test. After passing this convergence test, we generate $10^4$ more samples without updating the covariance matrix to obtain the final parameter distributions. Fluctuations that appear for more than $10^3$ iterations are a result of sampling values of $\lambda_A$ and $\lambda_B$ from the true distributions (cf.~Fig.~S\ref{fig:lambdaA_lambdaB_distr}).} 
\label{fig:bayes_convergence}
\end{figure}

To evaluating Eq.~\eqref{eq:metropolis_app}, we have to compute the least square error for the new proposed parameter set $\theta^\ast$ according to Eq.~\eqref{eq:error} to obtain the likelihood function value $P\left(D|\theta^\ast\right)$ based on Eq.~\eqref{eq:likelihood}. To prepare for a convergence test, we set the variance of our likelihood distribution to $\sigma^2=1/10$. Convergence is measured in terms of the Gelman-Rubin Test~\cite{gelman2013bayesian}. Therefore, we consider four independent Markov Chains. Every chain is initialized with a different random parameter set that is drawn from the corresponding uniform distributions. The posterior parameter distribution is considered to be converged if the variance between the chains is similar to the variance within the chain (Gelman-Rubin Test).
In Fig.~\ref{fig:bayes_convergence}, we illustrate the convergence behavior. We first let the four chains evolve for $10^4$ iterations to then apply the Gelman-Rubin Test. After reaching convergence, we generate $10^4$ more samples without updating the covariance matrix. These samples define the posterior parameter distribution.
\section{Influence of noise}
\label{sec:noise}
\begin{figure}
	\centering
	\includegraphics[width=0.49\textwidth]{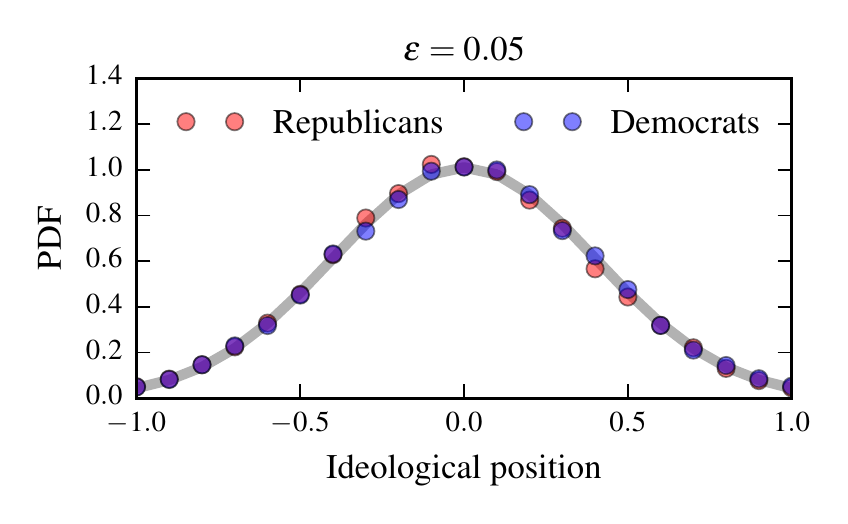}
	\includegraphics[width=0.49\textwidth]{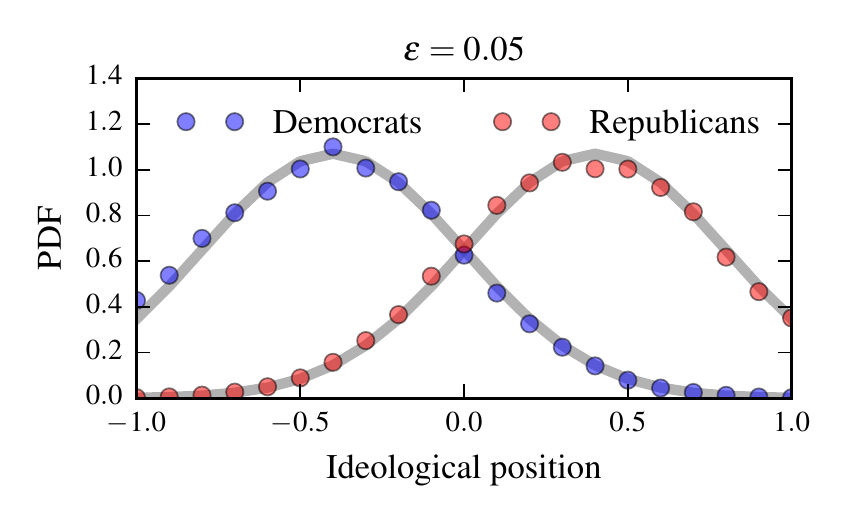}
	\includegraphics[width=0.49\textwidth]{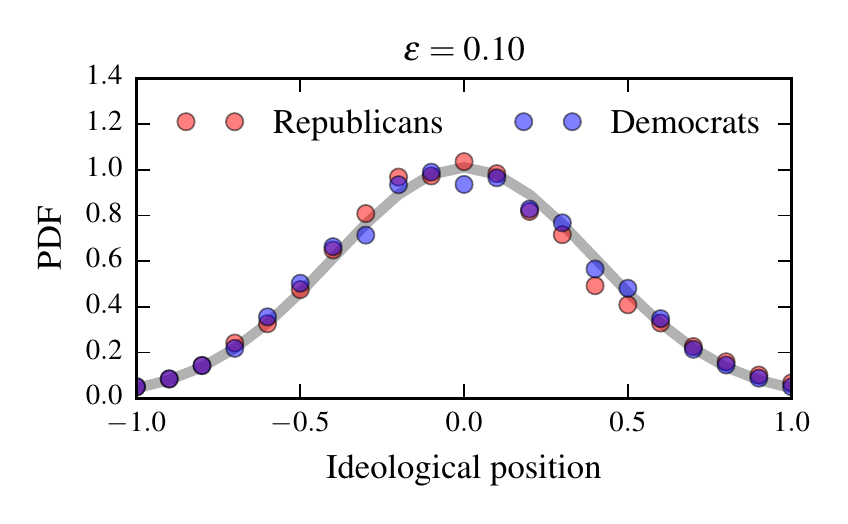}
	\includegraphics[width=0.49\textwidth]{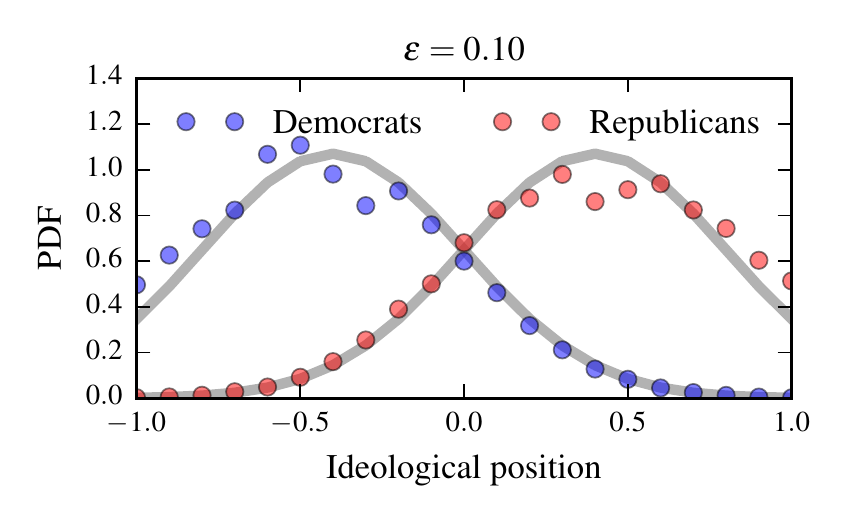}
	  \caption{
	  	\textbf{Influence of noise.} We show the influence of noise effects by considering relative fluctuations in the transition probabilities $p^A_i$, $q^A_i$, $p^B_i$, and $q^B_i$ ($i \in \{1,2,\dots,21\}$) by an amount of $\epsilon$. For example, we map each value of $p_i^A$ to $\left[1+\epsilon(2 u-1)\right] p_i^A$ where $u \sim \mathcal{U}(0,1)$. We apply the same mapping to the other transition probabilities. In the upper panels, we set $\epsilon=0.1$; and we set $\epsilon=0.1$ in the lower panels. The perturbed data points are represented by blue and red dots. The grey solid lines show unperturbed Gaussians. We rescaled the probabilities in the right panels according to Eqs.~(10) and (11) of the main text by setting $\lambda=\lambda_A=\lambda_B=1.13$.
  	 } 
	 \label{fig:noise}
\end{figure}
To examine the robustness of the transition-probability rescaling, we allow that the transition probabilities $p^A_i$, $q^A_i$, $p^B_i$, and $q^B_i$ of the discrete locations $i\in\{1,\dots,N\}$ may be subject to noise. To analyze the propagation of noise effects in the transition-probability rescaling process, we consider relative fluctuations in the initial transition probabilities by an amount of $\epsilon \in [0,1]$. For example, we map each value of $p_i^A$ to a value in the interval $\left[(1-\epsilon)p_i^A,(1+\epsilon)p_i^A\right]$. Specifically, let $u \sim \mathcal{U}(0,1)$ denote a uniformly distributed random variable. We then map $p^A_i$ to $\left[1+\epsilon(2 u-1)\right] p_i^A$. We apply the same mapping to the other transition probabilities. In Fig.~S\ref{fig:noise}, we show two examples of noise influences ($\epsilon=0.05$ and $\epsilon=0.1$) on the initial transition probabilities (left panels) and the rescaled ones (right panels). In both cases, the noisy data (blue and red dots) are still in good qualitative agreement with the unperturbed Gaussians (grey solid lines). This implies, that the transition-probability rescaling is also applicable in the case of noisy data if the noise influence is not too large. 
\newpage

%
%
%
%
\end{document}